\RequirePackage[l2tabu, orthodox]{nag}

\documentclass[10pt,conference]{IEEEtran}

\usepackage[T1]{fontenc}            
\usepackage[utf8]{inputenc}         
\usepackage{microtype}              
\usepackage[scaled=.8]{beramono}    
\usepackage[abbreviations]{foreign} 
\usepackage{balance}                
\usepackage{listings}
\usepackage[all]{nowidow}
\usepackage[dvipsnames]{xcolor}
\usepackage[
    colorlinks=true,
    allcolors=black
]{hyperref}
\usepackage{lipsum}
\usepackage{amsmath}
\usepackage{pifont}
\usepackage{booktabs}
\usepackage{graphicx}
\usepackage{subcaption}
\usepackage[numbers]{natbib}
\usepackage{cleveref}               

\definecolor{gusgreen}{RGB}{3, 125, 80}

\definecolor{keywordscolor}{RGB}{0, 51, 179}
\definecolor{stringcolor}{RGB}{6, 125, 23}
\definecolor{commentscolor}{RGB}{140, 140, 140}
\definecolor{annotationscolor}{RGB}{100, 100, 100}
\definecolor{lstbgcolor}{RGB}{251, 251, 251}

\lstdefinestyle{java}{
	language=Java,
    tabsize = 2, 
    showstringspaces = false, 
    numbers = none, 
    backgroundcolor = \color{lstbgcolor},
    commentstyle = \color{commentscolor}, 
    keywordstyle = \bfseries\color{keywordscolor}, 
    stringstyle = \color{stringcolor}, 
    rulecolor = \color{black}, 
    basicstyle = \footnotesize\ttfamily, 
    breaklines = true, 
    frame = lrtb,
    framexleftmargin = 3pt,
    xleftmargin = 3pt,
    escapeinside={!*}{*!},
    literate={->}{$\rightarrow$}{1}
}

\newcommand*{\ijava}[1]{\lstinline[basicstyle=\ttfamily,language=Java,keywordstyle=\mdseries\color{keywordscolor},morekeywords={sealed}]{#1}}
\newcommand*{\etclst}{\colorbox{blue!10}{\textbf{\vphantom{A}\dots}}}

\newcommand*{\gilesi}{\textsc{Gilesi}\@\xspace}

\newcommand*{\sembid}{\textsc{Sembid}\@\xspace}
\newcommand*{\updatera}{\textsc{Uppdatera}\@\xspace}
\newcommand*{\debbi}{\textsc{DeBBI}\@\xspace}
\newcommand*{\compcheck}{\textsc{CompCheck}\@\xspace}

\begin{document}

\title{Client--Library Compatibility Testing\\with API Interaction Snapshots}

 \author{\IEEEauthorblockN{Gustave Monce\IEEEauthorrefmark{1}, Thomas Degueule\IEEEauthorrefmark{1},
 	Jean-Rémy Falleri\IEEEauthorrefmark{1}\IEEEauthorrefmark{2} and Romain Robbes\IEEEauthorrefmark{1}}
 	\IEEEauthorblockA{Univ. Bordeaux, CNRS, Bordeaux INP, LaBRI, UMR 5800, F-33400 Talence, France\\
 	\IEEEauthorrefmark{1}firstname.lastname@labri.fr}
 	\IEEEauthorblockA{\IEEEauthorrefmark{2}Institut Universitaire de France}}


\maketitle

\begin{abstract}
Modern software development heavily relies on third-party libraries to speed up development and enhance quality.
As libraries evolve, they may break the tacit contract established with their clients by introducing behavioral breaking changes (BBCs) that alter run-time behavior and silently break client applications without being detected at compile time.
Traditional regression tests on the client side often fail to detect such BBCs, either due to limited library coverage or weak assertions that do not sufficiently exercise the library's expected behavior.

To address this issue, we propose a novel approach to client--library compatibility testing that leverages existing client tests in a novel way.
Instead of relying on developer-written assertions, we propose recording the actual interactions at the API boundary during the execution of client tests (protocol, input and output values, exceptions, \etc).
These sequences of API interactions are stored as \emph{snapshots} which capture the exact contract expected by a client at a specific point in time.
As the library evolves, we compare the original and new snapshots to identify perturbations in the contract, flag potential BBCs, and notify clients.

We implement this technique in our prototype tool \gilesi, a Java framework that automatically instruments library APIs, records snapshots, and compares them.
Through a preliminary case study on several client--library pairs with artificially seeded BBCs, we show that \gilesi reliably detects BBCs missed by client test suites.
\end{abstract}


\section{Introduction}

Modern software system development typically involves the use of third-party software libraries.
Instead of creating new systems from scratch, developers incorporate libraries with desired functionalities into their projects.
Application Programming Interfaces (APIs) govern the interactions between a library and its clients.
Whenever a library evolves, however, it may break the contract previously established with its clients by introducing breaking changes.
Breaking changes are either syntactic or semantic in nature~\cite{jayasuriya2025extended}.
Syntactic changes affect the API of a library, \ie the signatures of exported symbols accessed from client code (types, methods, \etc).
Semantic (or behavioral) changes affect the run-time behavior of a library and can alter client functionalities or trigger run-time errors.

These frictions at the boundary between clients and libraries have serious implications for both parties.
Clients must be warned of potential changes that could impact them and must protect themselves from vulnerabilities and supply chain attacks~\cite{ladisa23journey}.
Libraries must avoid carelessly breaking their clients to retain their trust and evolve gracefully~\cite{bogart21when}.
While there has been significant work on the prevalence~\cite{ochoa22breaking,jayasuriya2025extended}, detection~\cite{latappy25roseau,brito18apidiff}, and remediation~\cite{gao21apifix,zhong24compiler} of syntactic changes, behavioral breaking changes (BBCs) remain a challenge~\cite{zhang22has,mostafa17experience}.
Library tests---particularly regression tests---are well-suited to detect BBCs~\cite{liu23more}.
However, they are written by library maintainers and may not accurately reflect the actual usage of the library in the wild, which can differ significantly from the maintainers' expectations~\cite{schittekat22can,monce24lightweight}.

How can client projects ensure that changes introduced in a new library release will not affect them?
Client developers naturally write regression tests to validate their own logic, which indirectly exercise the libraries upon which their logic is built.
However, empirical evidence suggests that client tests are often too weak and ineffective in detecting BBCs introduced in library releases~\cite{jayasuriya2024understanding,gyori2018evaluating,hejderup22can}.
This can be caused by poor coverage of the library from the tests, weak assertions, and the significant distance between the tests and the exercised method in the library~\cite{niedermayr2019stack}.
Approaches in the literature that tackle this problem either do not take into account how libraries are used in client code~\cite{zhang2022has,chen20taming}, suffer from false positives~\cite{hejderup22can}, or adopt a crowd-sourced approach that assumes the existence of already migrated clients~\cite{chen20taming,zhu23client}.


In this paper, we propose a novel approach to leverage existing client tests and increase their ability to detect library BBCs.
We propose to infer the behavioral contract between a client and a library by observing the interactions (protocol and values exchanged) at the API boundary between the client and the library while client tests are running.
The sequence of interactions recorded during the execution of a test forms a \emph{snapshot} that embodies the precise expectations of a healthy client towards the library at a given point in time.
When the library is upgraded, the snapshots produced against the new library version can be compared against the original snapshots to detect differences in the interactions and report potential BBCs.
As snapshots are recorded directly at the API boundary, they are more sensitive to potential perturbations in the library's behavior and are more likely to catch BBCs than client tests.

We present \gilesi, a prototype implementation of our approach for Java libraries.\footnote{\url{https://github.com/alien-tools/gilesi}}
\gilesi seamlessly instruments software libraries to record interaction snapshots during the execution of client tests and compare them when the library is upgraded.
We showcase the capabilities of our approach and \gilesi on an exploratory case study involving the popular libraries \textsc{Jsoup} and \textsc{Commons-Lang3}, as well as 27 of their clients.
We find that \gilesi can detect artificial BBCs introduced in the libraries that are missed by client test suites.



\section{Background and Motivation}
\label{sec:background}

Application Programming Interfaces (APIs) comprise a set of publicly accessible symbols---such as types, methods, and fields---that external code can access and interact with~\cite{monce24lightweight}.
Clients, in turn, rely on these APIs to implement their own features, leveraging them by instantiating and extending types, accessing fields, invoking methods, \etc.

Only a subset of the API exposed by a library is typically used in a particular client~\cite{hejderup22can,harrand22api}.
This subset, the syntactic footprint~\cite{monce24lightweight}, corresponds to the specific elements of the API on which the client depends.
Breaking changes (BCs) in these elements can impact client code in diverse ways.
Syntactic BCs (SBCs) affect the signature of these elements and manifest as compilation or linking errors, while behavioral BCs (BBCs) affect their run-time behavior and manifest as changed program state and values, run-time exceptions, or crashes~\cite{mostafa17experience,reyes24bump,jayasuriya2024understanding}.
SBCs have been extensively studied in the literature~\cite{ochoa22breaking,jayasuriya2025extended} and can be detected automatically with good accuracy by static analysis tools~\cite{latappy25roseau}.
Comparatively, BBCs have received little attention and are particularly insidious as they cannot be reliably detected statically~\cite{zhang22has}.

\Cref{fig:bc} depicts an example BBC introduced in \texttt{common-text} and reported in \href{https://issues.apache.org/jira/browse/TEXT-219}{TEXT-219}.
The API method \texttt{StringTokenizer\#getTokenList} originally returns a list copy of an internal array of tokens.
In commit \href{https://github.com/apache/commons-text/commit/2d1ab7ea72298949900df47f65b4f71d56411f0b}{2d1ab7}, released with version \texttt{1.10}, a maintainer simplified various parts of the code and used the built-in utility \texttt{Arrays\#asList} to convert the array into a list.
This seemingly innocuous change introduced two adverse effects:~the returned list implementation is now fixed-size, and it is backed by the original array, meaning that changing either will change the other.
Thus, in contrast with the previous version, attempting to insert or remove an element in the returned list will raise a run-time exception, and reordering the list may have unintended side effects.

\begin{lstlisting}[float,style=java, caption={An example BBC introduced in \texttt{commons-text} commit \href{https://github.com/apache/commons-text/commit/2d1ab7ea72298949900df47f65b4f71d56411f0b}{2d1ab7}. The regression was identified in \href{https://issues.apache.org/jira/browse/TEXT-219}{TEXT-219} and later fixed in commit \href{https://github.com/apache/commons-text/commit/f9846b10b2365a36f95f63ff9f90e0f8847f901b}{f9846b}.}, label=fig:bc]
public class StringTokenizer !*\etclst*! {
  public List<String> getTokenList() { // API 1.9
    List<String> list = new ArrayList<>(tokens.length);
    Collections.addAll(list, tokens);
    return list;
    // In 1.10, the entire method was replaced with
    return Arrays.asList(tokens);
  } }
class Products { // Client code
	static List<String> fetchProducts(String products) {
		StringTokenizer st = new StringTokenizer(products);
		st.setDelimiterChar(',');
		return st.getTokenList();
	} }
class ProductsTest { // Client test
	@Test
	void test_fetchProducts() {
		String str = "apple,banana";
		List<String> fruits = Products.fetchProducts(str);
		assertEquals(2, fruits.size());
		assertThat(fruits, contains("apple", "banana"));
	} }
\end{lstlisting}

\subsection{Regression Testing}
To mitigate the risks associated with BBCs, clients naturally write tests to validate their own logic, which indirectly exercises the logic of the APIs they use.
In particular, developers use \emph{regression tests} to ensure that changes do not unexpectedly alter their software's behavior~\cite{gyori2018evaluating,liu23more}.
When libraries are updated to newer versions, these tests can sometimes catch BBCs introduced in the library and inform client developers~\cite{hejderup22can,jayasuriya2024understanding}.
Executing the clients' test suites is also beneficial to library maintainers, as it can exercise their library in ways that they have not necessarily anticipated and catch regressions unseen by their own regression tests~\cite{gyori2018evaluating}.
A popular implementation of this idea is GitHub's \textsc{Dependabot}, which adopts a crowd-sourced approach to run the CI and tests of client projects that have migrated to a new version and identify potentially unsafe upgrades.

But \emph{are client tests really effective in detecting BBCs in third-party libraries?}
In a large corpus of 8,086 Maven artifacts with passing test suites, \citeauthor{jayasuriya2024understanding} find that upgrading their dependencies to the latest available version triggers test failures in only 2.3\% of cases~\cite{jayasuriya2024understanding}.
\citeauthor{gyori2018evaluating} find that 26.9\% of libraries that can be updated in client projects result in test failures~\cite{gyori2018evaluating}.
\citeauthor{hejderup22can} analyze 262 Java projects and find that client tests detect only 47\% (resp.\,35\%) of artificial faults seeded in direct (resp.\,indirect) dependencies~\cite{hejderup22can}.
These empirical results suggest that existing client test suites are often too weak and ineffective in detecting BBCs in libraries.

Regression tests can fail to detect BBCs for two reasons:~either they lack coverage and do not exercise the execution paths affected by BBCs, or their assertions are too weak to observe the change.
Indeed, for BBCs to be detected, they must manifest at the API's boundary and propagate through client code up to a strong enough assertion that can observe the altered state.
At each step, as the distance to change increases, lenient code can alter the observability of the change and reduce the likelihood of a failing assertion~\cite{niedermayr2019stack}.
For instance, exception-swallowing code can prevent a newly raised exception from being detected by tests, or return values can simply be ignored.
Subtle behavioral changes are particularly unlikely to be detected in this manner, leaving clients unaware of potential errors that may occur at run time.
In \Cref{fig:bc}, although client tests accurately assert the behavior of the code relying on \texttt{StringTokenizer\#getTokenList}, they do not detect the regression.
Other parts of the client that would attempt to alter the returned list might break as a result, and the developers would not be informed before the issue manifests at run time.

Interestingly, \citeauthor{jayasuriya2024understanding} note
that, even when BBCs do trigger test failures, the distance between the root cause of the error and the failing assertion severely hurts diagnosis and remediation~\cite{jayasuriya2024understanding}.
To alleviate this issue, developers sometimes write tests specifically to detect changes and regressions in the behavior of the libraries they use.
These tests directly exercise the library’s APIs---thus reducing distance and increasing observability---and focus on verifying the expected outputs and side effects.
However, client tests are primarily concerned with validating their own logic, and such library-focused tests are relatively rare in practice.

\subsection{Dedicated Approaches}
Several approaches have attempted to alleviate these issues and to detect BBCs in software libraries more reliably.
They can be classified along three axes:~(i)~whether they consider the entire API of a library or only the subset used by a particular client, (ii)~whether they exploit crowd-sourced knowledge from other clients or can be applied to a single client immediately after a release, and (iii)~whether they rely on test execution or static analysis of library changes.

\sembid and \debbi are two approaches aiming at detecting BBCs introduced in new library versions.
\sembid employs static analysis to compute semantic diffs between two API method versions and flag changes as potential BBCs~\cite{zhang2022has}.
\debbi, on the other hand, adopts a crowd-sourced approach to aggregate the test suites of different clients using the same library~\cite{chen20taming}.
As executing the resulting tests can be costly, \debbi uses information retrieval techniques to prioritize the execution of test cases that are more likely to detect regressions in a given upgrade.
While these approaches are beneficial for library maintainers who wish to avoid breaking any of their clients, they do not account for the specificity of particular clients that may tolerate BBCs that do not impact them.

\updatera and \compcheck are two approaches aiming at detecting BBCs affecting a particular client.
\updatera is a client-specific approach that classifies changes in the methods reached by a particular client as regressions by identifying suspicious changes in control flow and dataflow using structural diffs between the two versions~\cite{hejderup22can}.
\compcheck is another client-specific approach that leverages crowd-sourced knowledge to generate incompatibility-revealing tests adapted from other clients using the library in a similar way~\cite{zhu23client}.
While \updatera sometimes produces false positives since the diff matching rules are too strict, \compcheck is guaranteed to produce true positives only with an accompanying test case.
However, it may still suffer from false negatives when the generated assertions are too weak and not sensitive enough to detect the difference.


\section{Approach}
\label{sec:approach}

\begin{figure}
	\centering
	\includegraphics[width=.6\linewidth]{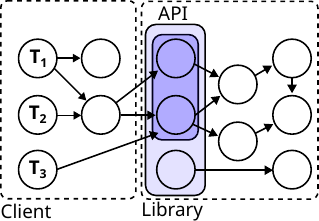}
	\caption{Libraries expose their features through dedicated APIs. Clients use a subset of these APIs to implement their own features. For the sake of readability, only methods are depicted.}
	\label{fig:overview}
\end{figure}


Our hypothesis is that developers write test cases to exercise important features and execution paths in their code.
When the implementation of these features relies on third-party code, however, developer tests may not be able to properly assert its behavior due to the distance between the test case and the method it tests~\cite{niedermayr2019stack} and the difficulty of writing assertions on code they do not own.

Our intuition is that BBCs introduced in libraries primarily manifest as perturbations in the sequence of invocations (protocol) or values exchanged at the API boundary before and after the change (\Cref{fig:overview}).
These include return values, in-out parameters, exceptions, or crashes.
These perturbations reflect deviations from the library's expected behavior but may not always propagate far enough through client code to trigger existing assertions.
Nevertheless, they can still affect the client's behavior and should be detected.

\subsection{Overview}
To address this challenge, we propose an approach that involves instrumenting and recording interactions at the API boundary during the execution of client tests.
When running a client's test suite against a specific version of the library, we capture these interactions and record them as \emph{snapshots}, \ie concrete traces that materialize the exact expectations of the client towards the library at a given time.
Inspired by snapshot testing~\cite{cruz23snapshot} and production monitoring techniques such as \textsc{pankti}~\cite{tiwari_production_2022}, snapshots document the behavioral contract between a client and a library.

More precisely, each test case produces a snapshot $S$ represented as a sequence of API interactions.
An interaction $I = \langle m, o, \langle p_1, \dots, p_n \rangle, r \rangle$ records, for a particular API method $m$, the receiver object $o$ on which the method is invoked, the list of input values $p_i$ passed as arguments and the (typed) result of its execution $r$---either a concrete value or an exception.
For example, in \Cref{fig:bc}, the initial execution of the test first invokes the client code depicted, which in turn invokes the API.
The resulting snapshot for this test consists of a single API interaction $S_1 = (\langle \texttt{getTokenList}, o_1, \emptyset, \texttt{ArrayList}(apple, banana) \rangle)$.

When the library is upgraded, we run the client's test suite again and collect a new set of snapshots.
The new snapshots can be compared to those recorded with the previous version to identify BBCs.
Any difference between the original and the new snapshots indicates a change in the library's behavior.
These discrepancies may arise from altered usage protocols (when the order of method invocations differs) or different return values (when an API produces a different or exceptional result for the same inputs on the same object).
Since assertions are now evaluated at the API boundary, our approach is more sensitive to subtle behavioral changes that may not cause existing tests to fail, but still affect client logic.
This enables clients to detect and review behavioral changes, even when those changes do not propagate to existing test assertions.
In \Cref{fig:bc}, when the API method is updated, the new snapshot consists of a single API interaction $S_2 = \langle \texttt{getTokenList}, o_1, \emptyset, \texttt{Arrays\$ArrayList}(apple, banana) \rangle$.
Because $S_1 \neq S_2$, the two snapshots differ, and a BBC is reported for the method \texttt{getTokenList} in the new version.

Interestingly, our approach addresses three challenges typically faced in the detection of BBCs.
First, it naturally captures reflective calls, as all method invocations are logged regardless of how they are dispatched.
Second, it can detect changes in transitive dependencies as long as they propagate up to the API of a direct dependency to impact client code.
More generally, it does not matter where the change is introduced as long as it propagates and can be observed at the API boundary manipulated by the client.
Finally, snapshots explicitly document the exact behavior of the two library versions, making it easier to understand and debug BBCs.

\subsection{Instrumenting the API}
As shown in \Cref{fig:overview}, only a subset of library code is typically exposed to clients through a dedicated API.
In object-oriented languages like Java, this typically includes exported types, methods, and fields.
BBCs, whether introduced in public or internal code, must eventually surface at the API boundary.
Thus, a first step is to delimit the API of the library and the subset used in a particular client.

To target only the relevant API used by a given client, we extract a client-specific syntactic usage footprint using UCov~\cite{monce24lightweight}, a lightweight static analysis tool.
This enables us to focus on API symbols that can affect the client.
These API symbols are then instrumented to log the values passed into and returned from each method during test execution.
We use Java’s instrumentation APIs in conjunction with Byte Buddy to implement a run-time agent that intercepts class loading events, identifies relevant method signatures (as determined by UCov), and injects probes at their entry and exit points.
These probes are responsible for capturing the receiver object, input arguments, and return values.
Aside from the added probes, the method's behavior remains unchanged.

\subsection{Recording \& Comparing Snapshots}
Once instrumentation is in place, running the client's test suite triggers the probes on each direct or indirect API call.
The main challenge lies in serializing the observed values into stable snapshots that can be compared across library versions.

While serialization is straightforward for standard JDK types (\eg primitives, collections), it becomes difficult for deeply nested or library-defined objects.
We use the \textsc{XStream} framework to serialize standard types.
For complex objects, we defer serialization until additional method calls on them expose simpler, serializable data.
For example, when the library returns a library-defined object, we record the interaction but only keep track of the object's identifier without attempting to serialize it.
Then, we wait until other API methods are invoked on it and record those as well.
The intuition here is that when the new API returns a different complex object, it cannot affect the client until concrete services are invoked.

Comparing snapshots first involves verifying that the ordering of interactions is identical between the old and new snapshots.
When this is the case, interactions are compared pair-wise to verify that input and output values are the same.
To match interactions, objects are given a unique identifier that allows tracing method invocations on the same objects throughout the execution.







\section{Case Study}
\label{sec:eval}

In this section, we report on a preliminary case study of \gilesi's ability to detect BBCs compared to client test suites.
Our case study involves the Java libraries \textsc{Jsoup} and $6$ of its clients as well as \textsc{Commons-Lang3} and $21$ of its clients.
The data and notebooks discussed in this paper are available on Zenodo (\url{https://doi.org/10.5281/zenodo.16411967}).

\subsection{Protocol}

As an initial sanity check, we first run the test suites of client projects against the original version of the libraries twice.
Then, we filter out any client with failing or flaky test suites.
We also record and store which API methods are reached during the execution of the tests.

To simulate the introduction of BBCs in these libraries, we implement another simple Java agent that dynamically introduces extreme mutations in API methods of interest.
The agent removes all code in the original implementation and replaces it with a single return statement that returns a default value (\texttt{null} for reference types, zero for integers, \etc).
This simulates a drastic change in behavior that should normally be caught by both client test suites and \gilesi.

Then, we evaluate the ability of client test suites and \gilesi to detect the introduced BBCs using a simple mutation score.
First, we use \gilesi to record the set of snapshots produced when running the client test suites against the original version of the libraries.
We repeat this operation twice to ensure that \gilesi produces stable snapshots and discard any test that yields flaky snapshots.
Then, for each API method reached in client tests, we use the agent to insert the extreme mutations, one at a time, and run the client test suites another time.
If any of the tests fail after introducing the mutation, then we consider that the client test suite successfully kills the mutant and detects the BBC.
If any of the snapshots extracted after the mutation is introduced differ from the corresponding snapshot extracted on the original version of the library, then we consider that \gilesi successfully detects the BBC and kills the mutant.

\subsection{Overall results}

\Cref{tab:aggmutscores} presents the mutation scores obtained by client tests and \gilesi following our protocol.
Overall, \gilesi detects 96\% of the 158 mutants introduced in both libraries, while client test suites only detect 89\% of them.
All mutants killed by client tests are also killed by \gilesi.
Conversely, \gilesi detects 10 extreme mutations missed by client tests.

However, there are still seven mutants that are not detected by \gilesi.
These cases stem from mutations that alter library-side I/O side effects without changing return values (\eg on \ijava{void} methods), or produce equivalent behavior because the original methods already return default values to all clients.

In this experiment, we rely on extreme mutations to simulate the introduction of BBCs.
We hypothesize that finer-grained, classical mutation operators would be less likely to trigger test failures on the client side, while \gilesi should retain much of its detection abilities, being closer to the seeded fault.
Future experiments should thus further investigate how client tests and \gilesi handle finer-grained mutations and real BBCs.

\begin{table}[tb]
	\small
    \caption{Aggregated mutation scores.}
\begin{tabular}{p{0.35\linewidth}rrr}
        \textbf{Library} & \textbf{Mutants} & \multicolumn{2}{c}{\textbf{Killed by}}\\
        \cmidrule(lr){3-4}
        & & Tests & \gilesi \\
        \midrule
        \textsc{commons-lang3} & 106 & 100 & 105\\
        \textsc{jsoup} & 52	& 41 & 46\\
    \end{tabular}
    \label{tab:aggmutscores}
\end{table}

\subsection{Illustrative Cases}
As part of our case study, we review two noteworthy cases that highlight how \gilesi can help client developers protect themselves against library BBCs, including for clients with strong test suites.

\subsubsection{Case 1: \textsc{devops-comdor} and \textsc{commons-lang3}}

\textsc{devops-comdor} features functionality designed to help create issues on the code hosting platform GitHub.
Part of the test suite of \textsc{devops-comdor} aims to test the creation of an issue when an exception is caught and to check whether the created issue contains correct information about such an exception.
Information about the exception is extracted using the \texttt{getStackTrace} method of \textsc{Commons-Lang3} (\Cref{lst:case1}).

With the original version of the library, the issue is successfully created with the correct title and body.
When running the same test on the mutated library, \texttt{getStackTrace} no longer returns information about the exception.
Thus, the created issue lacks crucial information.
However, the corresponding test only asserts parts of the issue body and misses the introduced BBC.
The client is thus vulnerable to future evolutions of \textsc{Commons-lang3} and may not detect potential regressions.

On the other hand, \gilesi manages to catch the regression automatically.
The original snapshot for this test identifies that \texttt{getStackTrace} receives an \texttt{IOException} and produces a string describing the exception.
The new snapshot produced after the mutation is introduced maps the exact same input to a different output, a \texttt{null} value.
As the two snapshots differ, the BBC is detected and successfully reported to the user.

\begin{lstlisting}[float,style=java,caption={API, client code, and client test showcasing a missed BBC in \textsc{devops-comdor}}, label=lst:case1]
public class ExceptionUtils { // API
  // Input (IOException): <message>expected</message>
  //                      <stackTrace> <trace>co.comdor.!*\etclst*!
  public static String getStackTrace(Throwable throwable) {
    StringWriter sw = new(); PrintWriter pw = new(sw, true);
    throwable.printStackTrace(pw);
    return sw.getBuffer().toString();
  }
  // Original output: "IOException: expected at Vigilant !*\etclst*!
  // Mutant   output: null
}
public class VigilantAction { // Client
  public void perform() throws IOException {
    try { og.perform(); }
    catch (IOException | RuntimeException ex) {
      String title = "!*\etclst*!";
      String body = "!*\etclst*! Exception:" +
        ExceptionUtils.getStackTrace(ex);
      Issue created = gh.repos().get("amihaiemil/comdor")
        .issues().create(title, body); }
  } }
public class VigilantActionTestCase { // Client test
  @Test public void opensIssueOnIOException() throws Exception {
    Github gh = mockGithub(); Action og = mockAction();
    doThrow(new IOException("expected")).when(og).perform();
    Action vigilant = new(og, gh); vigilant.perform();
    Issues all = gh.repos().get("amihaiemil/comdor").issues();
    assertThat(all, iterableWithSize(1));
    Issue op = all.get(1);
    String body = "!*\etclst*! Exception:" +
      ExceptionUtils.getStackTrace(ex);
    assertThat(op.json().getString("body"), startsWith(body));
  } }
\end{lstlisting}

\subsubsection{Case 2: \textsc{chyxion-table-to-xls} and \textsc{JSoup}}

\textsc{chyxion-table-to-xls} features functionality designed to turn an HTML document containing a table into an XLS spreadsheet.
This client leverages several \textsc{JSoup} features for parsing the HTML document, including \texttt{attr} to retrieve the string values of attributes.

In this example, the test verifies that a predefined HTML document can be successfully converted into a spreadsheet.
Running the test against the original version of the library successfully creates a new spreadsheet with appropriate rows and columns.
Running the test against the mutated version of the library, however, returns a spreadsheet with missing rows and columns.
Indeed, the test merely verifies that the conversion runs successfully without crashes or exceptions, without asserting the shape of the returned spreadsheet.
Therefore, the client is vulnerable to potential regressions in the evolution of \textsc{JSoup}.

On the other hand, \gilesi manages to catch the regression automatically.
The snapshots produced with the original and mutated versions differ on multiple aspects.
The first interaction with the \texttt{attr} method produces a different outcome, and the number of interactions differs.
\gilesi successfully catches the difference that can be reported to the client.

\section{Conclusion \& Research Roadmap}
\label{sec:discussion}

In this paper, we introduced a novel approach for detecting BBCs in software libraries by capturing API interaction snapshots during client test execution and comparing these snapshots across releases.
Through a case study involving real-world libraries and clients, we showed that \gilesi can identify BBCs missed by client test suites.

To realize the full potential of \gilesi, future efforts should scale its evaluation across a broader set of libraries and involve finer-grained mutation operators or real releases, as well as account for the presence of nondeterminism, concurrency, and side effects.
Addressing these limitations in future work will be critical to improving detection accuracy and enabling practical use by software developers.

\balance
\bibliographystyle{IEEEtranN}
\bibliography{IEEEabrv,main}

\end{document}